\documentclass[12pt]{iopart}

\usepackage{graphicx}
\usepackage{iopams}

\newcommand \be{\begin{eqnarray}}
\newcommand \ee{\end{eqnarray}}
\newcommand \ba{\begin{eqnarray}}
\newcommand \ea{\end{eqnarray}}

\begin{document}

\title[Positive and negative streamers in ambient air: measurements]
{Positive and negative streamers in ambient air: measuring diameter, velocity and dissipated energy}

\author{T M P Briels$^1$, J Kos$^1$, G J J Winands$^2$, E M van Veldhuizen$^1$ and U Ebert$^{1,3}$}

\address{$^1$ Department of Applied Physics and $^2$ Department of Electrical Engineering, Technische
Universiteit Eindhoven, P O Box 513, 5600 MB Eindhoven, The Netherlands,\\
$^3$ Centrum voor Wiskunde en Informatica (CWI),
P O Box 94079, 1090 GB Amsterdam, The Netherlands}

\ead{e.m.v.veldhuizen@tue.nl, ebert@cwi.nl}

\begin{abstract}
Positive and negative streamers are studied in ambient air at 1 bar; they emerge from a needle electrode
placed 40 mm above a planar electrode. The amplitudes of the applied voltage pulses range from 5 to 96 kV;
most pulses have rise times of 30 ns or shorter. Diameters, velocities and energies of the streamers are
measured. Two regimes are identified; a low voltage regime where only positive streamers appear and a high
voltage regime where both positive and negative streamers exist. Below 5 kV, no streamers emerge. In the
range from 5 to 40 kV, positive streamers form, while the negative discharges only form a glowing cloud at
the electrode tip, but no streamers. For 5 to 20 kV, diameters and velocities of the positive streamers have
the minimal values of $d=0.2$ mm and $v\approx10^5$ m/s. For 20 to 40 kV, their diameters increase by a
factor 6 while the voltage increases only by a factor 2. Above the transition value of 40 kV, streamers of
both polarities form; they strongly resemble each other, though the positive ones propagate further; their
diameters continue to increase with applied voltage. For 96 kV, positive streamers attain diameters of 3 mm
and velocities of $4\cdot10^6$ m/s, negative streamers are about 20\% slower and thinner. An empirical fit
formula for the relation between velocity $v$ and diameter $d$ is $v=0.5~d^2/$(mm~ns) for both polarities.
Streamers of both polarities dissipate energies of the order of several mJ per streamer while crossing the
gap.
\end{abstract}


\submitto{The paper is accepted for publication in the cluster issue on ``Streamers, Sprites and Lightning''
in \JPD}

\maketitle

\section{Introduction}

Streamers of both polarities appear in many phenomena in nature~\cite{rai91,Baz,wil06}. Industrial
applications, on the other hand, have largely focussed on positive (cathode directed) streamers
\cite{wag66,PG76,Mar75,Bas97,Spy89,Taj89,Crey94,All95,Ale96,Pan00,Pan01,Pan03,Pan04,panchpre,Pan05,OO03,OO05,OO08}
as they are easier to create around sharp tips than negative (anode directed) ones~\cite{rai91}. This even
has lead to a tendency in the applied physics and electrical engineering literature to use the term
"streamer" as equivalent to "positive streamer". Positive streamer coronas are used in gas cleaning and
ozone production since they were long thought to have a higher energy efficiency than negative streamer
coronas~\cite{eddieboek}. However, recent investigations with a new generation of pulsed power supplies have
shown that negative streamers at voltages of 50 to 80~kV can convert pulsed electric energy into ozone with
an unequalled efficiency of 100 g/kWh in ambient air~\cite{win07,heesch08}. Furthermore, negative DC-corona
is used in dust precipitators to charge small particles that then can be drawn out of a gas stream by an
electric field. The properties of these negative streamers at voltages above 50~kV largely resemble those of
positive ones~\cite{win07,heesch08,BlomArt,BlomPhD,hansieee,hansjpd}. A qualitative similarity between
positive and negative streamers in nitrogen with varying oxygen concentration is also found in experiments
in a 13 cm gap in protrusion-plane electrode geometry at voltages of 82 to 125~kV~\cite{yi02} while other
authors emphasize their difference~\cite{vel02}.

Up to now, studies have explored limited parameter regimes, seeing either positive streamers only, or both
positive and negative streamers. Here we present a systematic study over a wide parameter regime where
actually two regimes are seen, one with positive streamers only and another one with positive and negative
streamers. We investigate streamers in ambient laboratory air between a needle and a planar electrode at a
distance of 4 cm. We investigate a voltage range from 5 to 96~kV with three different voltage-supplies that
were previously used either in the studies of positive streamers at the physics
department~\cite{bri04,bri06,bri08,briieee,briscaling,nijdam}, or for both positive and negative streamers
at the electrical engineering department~\cite{win07,heesch08,BlomArt,BlomPhD,hansieee,hansjpd} of Eindhoven
University of Technology. We attribute similarity or major differences between positive and negative
streamers to different voltage ranges. For low voltages, thin positive streamers ignite and propagate
easily, while negative discharges require much higher ignition voltages, they form thick and short streamers
if at all. For voltages above 60~kV, positive and negative streamers become more and more alike.

Theoretical investigations of the difference between positive and negative streamers in three spatial
dimensions (using the cylindrical symmetry of a streamer to calculate effectively in the two dimensions $r$
and $z$) and including the photo-ionization effect in air are quite
rare~\cite{bab97,LiuP04,Luque2007/ApPhL,Luque2007/arXiv}, a thorough discussion and new results that closely
correspond to the experimental results of the present paper are presented in~\cite{PMTheo}.

For readers from the geophysical community, a short reminder is in place why we discuss the dependence on
applied voltage rather than on applied field. To create streamers efficiently, and to photograph them with
high spatial and temporal resolution, streamers in experiments and applications are typically emitted from
needle or wire electrodes that create strong electric fields in their neighborhood.\footnote{We remark that
similar mechanisms can take place next to charged droplets or ice particles in thunderclouds.} As a short
consideration --- e.g., of the example for a charged sphere --- shows, the electric field close to the
strongly curved electrode is mostly determined by voltage and electrode geometry and rather independent of
the distance to some distant grounded electrode. It is therefore physically evident and has been confirmed
by experiments~\cite{bri04,bri06}, that streamer inception and initial propagation is determined by applied
voltage and electrode geometry, and not by some hypothetical average field within the complete discharge
gap. The streamers start in a high field region and consecutively expand into a region with decreasing
field.

All streamers presented in this paper are created in a 40 mm gap in a needle to plane electrode geometry in
ambient air at atmospheric pressure. We have chosen to investigate the streamers in this gap length because
in longer gaps the streamers can branch and form thinner ones~\cite{bri06}, and at high voltages in
wire-plane electrode geometries they also have been seen to become thicker while propagating~\cite{hansjpd};
in smaller gaps the streamers have not enough space to develop completely~\cite{bri08,briscaling}, and it is
difficult to measure their velocities.

In~\cite{briscaling} we concentrated on the investigation of the thinnest positive streamers at varying
pressure and gas composition and used voltage supplies with ``slow'' rise times. In contrast, we here use
three voltage supplies with fast rise time to create the thickest streamers possible at the particular peak
voltage. The three supplies together cover the voltage range from 5 to 96~kV, their ranges overlap and we
find the resulting streamer properties to depend continuously on the voltage, independently of the used
supply. We also perform control experiments with one slower voltage supply to further confirm the statement
in~\cite{bri06} that power supplies will create similar streamer patterns only if their voltage rise time,
peak voltage and internal resistance are similar.

The paper is organized as follows. The experimental setups are described in section~\ref{setup}. The
evaluation procedure of photographs and the results are described in section~\ref{res}; the section is
ordered into a general overview, dependence on the voltage supply, stability field, streamer diameters,
streamer velocities, an empirical relation between diameter and velocity, and current and energy. Section
\ref{concl} contains summary and conclusions.

\section{Experimental setup}\label{setup}


\subsection{Voltage pulse generation}\label{voltagepulse}

\begin{figure}
\begin{center}
a) \includegraphics[width=10cm]{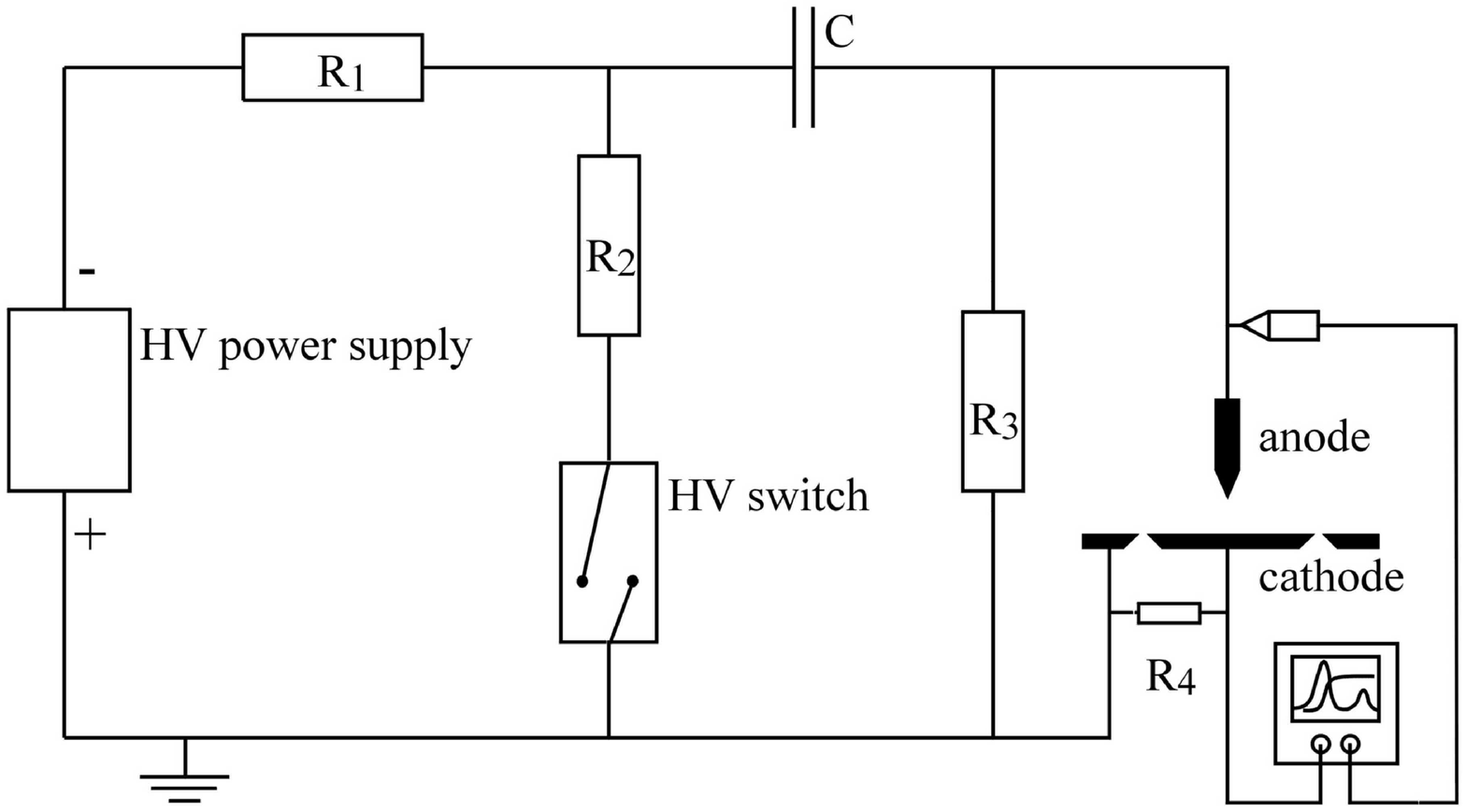}\\~\\~\\
b) \includegraphics[width=10cm]{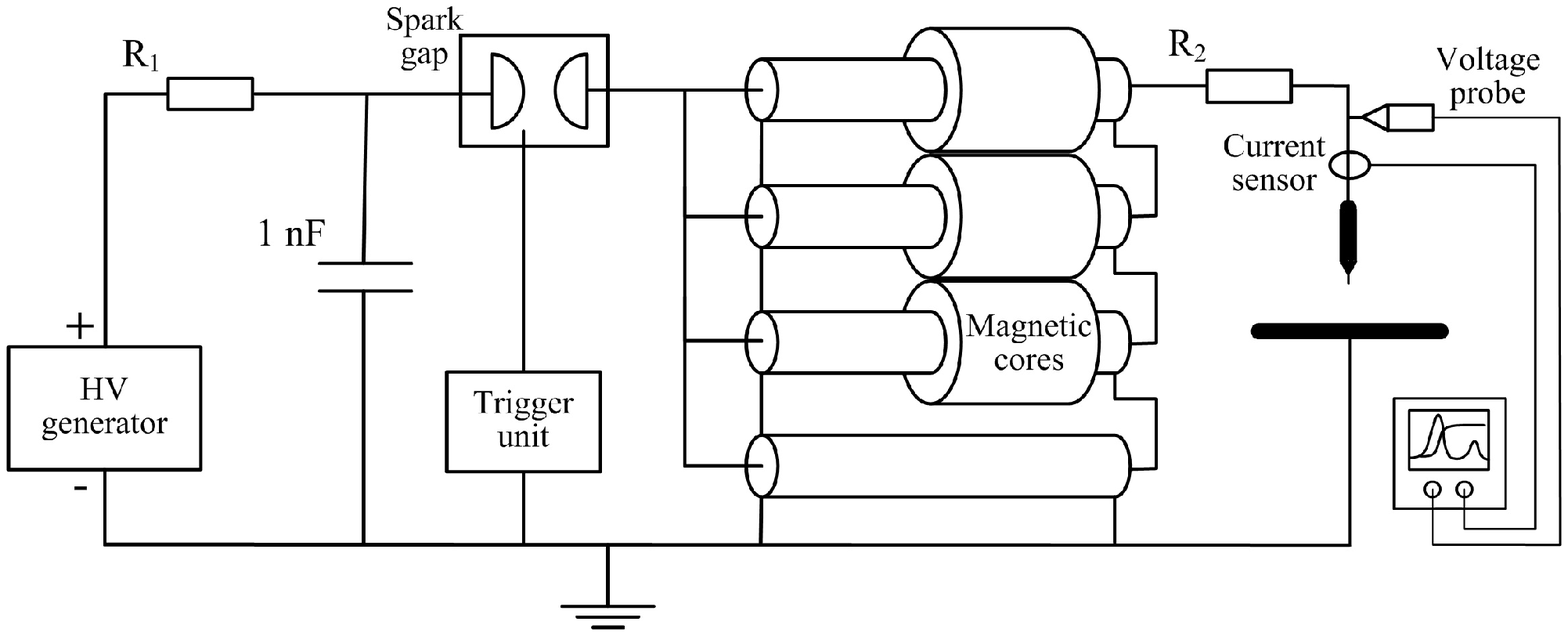}\\~\\~\\
c) \includegraphics[width=10cm]{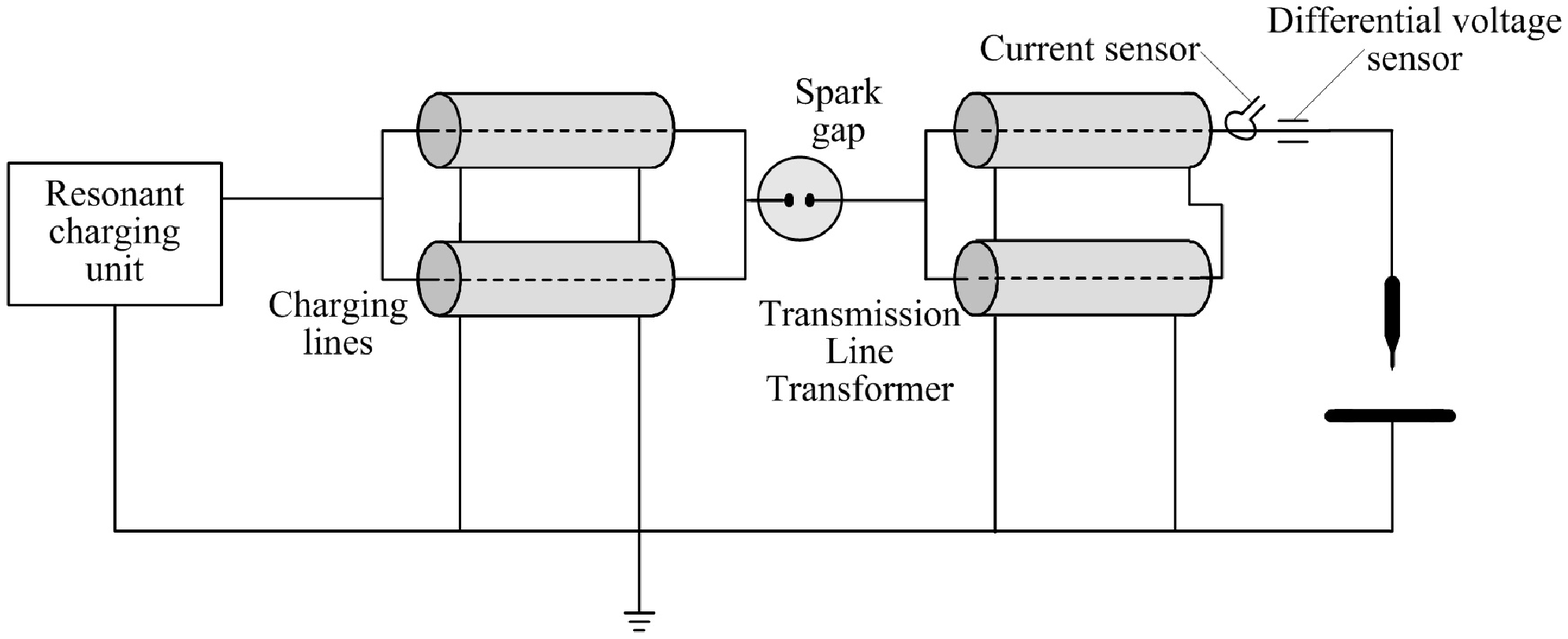}
\caption{\em a) C-supply, b) TLT-supply and c) PM-supply. C-supply and TLT-supply were used in the
experiments at the physics department and already described in \cite{bri06}; a detailed drawing of the
discharge chamber in these supplies including the current and voltage measurements is provided in Fig.~2
of~\cite{briscaling}. The PM-supply was used in the experiments at the electrical engineering department and
already discussed in \cite{win06}; the panel indicates the location of current and voltage measurements.
Different diagnostics was used at the two locations as described in the text.} \label{figsupplies}
\end{center}
\end{figure}

A pulsed power supply consists of three parts: (i)~charge storage, (ii)~switch, and (iii)~load. (i)~The
charge is conventionally stored in a capacitor, these are readily available for voltages up to 100 kV.
(ii)~The switch is the crucial part when rise times in the nanosecond range are required. The prevailing
choice still is the spark gap, modern versions combine robustness, high efficiency and long life
time~\cite{win05,liu06}. Alternatives such as semiconductors, whether or not in combination with magnetic
compression, offer in principle longer life time, but are vulnerable, have lower efficiency and are very
expensive. (iii)~The load is basically the corona discharge, but impedance matching is a major issue for
large systems \cite{yan01}. The three supplies used in the current study are sketched in
Fig.~\ref{figsupplies}; their properties are listed in table~\ref{tabsupplies}. In general, the voltage
pulse duration is kept short since long pulses at high voltages may allow the initial streamer to develop
into a much brighter arc or spark discharge which forms a risk for overexposure of the intensifier of the
CCD camera.

The simplest supply is called the C-supply; it was thoroughly described in \cite{bri06} and it is sketched
in Fig.~\ref{figsupplies}(a). In a C-supply, a capacitor first is charged and subsequently discharged over
the needle plate gap when the spark gap is closed. It gives the pulse exponential rise and decay times that
can be adapted with resistors $R_1$ and $R_2$. Therefore this supply is very versatile and easy to build.
Two versions of the C-supply were used as listed in table \ref{tabsupplies}. The C$_1$-supply has the
advantage of a short rise time $T_{rise}=30$~ns and quite short duration $T_{duration}=1~\mu$s, but this
limits the pulse amplitude. The C$_2$-supply has a longer rise time $T_{rise}=150$~ns and a longer pulse
duration $T_{duration}=1$~ms. It is used to confirm that the streamer structure changes when the rise time
becomes too long~\cite{bri06}; the long pulse duration might allow the transition to arc or spark. The
C-supplies deliver up to 60 kV.

\begin{table}
\begin{tabular}{c||c|c|c|c|c||c|c|c|c|}
name & $R_1$ & $R_2$ & $R_3$ & $C$ & $T_{\rm rise}$ & $T_{\rm duration}$ & range of $U$
    & polarity \\
    & (k$\Omega$) & (M$\Omega$)   & ($\Omega$)& (nF) & (ns) & ($\mu$s) & (kV)  & \\
\hline \hline
C$_1$       & 0         & 0.004  & 2.75    & 0.250 & 30   & 1 & 1 -- 60 & $+$         \\
\hline
C$_2$       & 2         & 1      & 2.75    & 1     & 150  & 1000 & 1 -- 60 & $+$ and $-$  \\ 
\hline
TLT         & 1         & ---      & ---     & 1     & 25   & 0.05 & 30 -- 60 & $+$         \\ 
\hline
PM          & ---      & ---     & ---     & 2 (PFL)  & 15  & 0.1  & 40 -- 96 & $+$ and $-$   \\ 
\hline
\end{tabular}
\caption{\em Characterization of the electric power supplies in the present experiments. The resistor $R_1$
influences the rise time of the pulse. The combination of $R_2$ and C determines the decay time of the
voltage pulse. $R_3$ is a series resistance to determine the discharge current. }\label{tabsupplies}
\end{table}

The pulse amplitude can be increased through a transmission line transformer (TLT) as shown in
Fig.~\ref{figsupplies}(b). In a TLT several coaxial cables are connected in series or in parallel at the
input or the output side in such a way that either voltage or current is multiplied. A TLT can also be used
as impedance matching network between the power source and the corona discharge reactor \cite{yan01}. The
second advantage is that  the pulse width and the decay time are very short. This system is further referred
to as TLT-supply, it was thoroughly described in \cite{bri06} as well. Its main characteristics are also
listed in table \ref{tabsupplies}; the voltage rise time is 25~ns, and the pulse duration is 50 ns. The
voltage range is 30 to 60 kV.

The third supply charges and discharges using coaxial cables. This supply is called the Power Modulator (PM)
supply and will be abbreviated as PM-supply, it is shown in Fig.~\ref{figsupplies}(c). Its output pulses
have the same characteristics as the TLT-supply. For charging it uses a pulse forming line (PFL) which leads
to a very high wall-plug efficiency (95\%). The supply is discussed in more detail in \cite{win07,win06}.
The rise time of the PM-supply is shorter than that of the TLT-supply, namely 15~ns, probably due to the
lower inductance of the load; the pulse duration is 100~ns. This supply has a range from 40 to 96 kV.

\subsection{Electrical measurements}

With the C-supplies, the voltage is measured by a resistive-capacitive divider (Tektronix P6015 for C$_1$ or
Northstar PVM4 for C$_2$) at the positions indicated in Fig.~2 of~\cite{briscaling}. The current through the
gap is measured with a Pearson current monitor (6585) for C$_1$ or via the voltage across a small series
resistor $R_s = 2.75 \Omega$ between cathode and ground for C$_2$. The outer ring across the cathode ensures
a well defined, low stray capacity and therefore a fast rise time of the current measurement \cite{gra87}.

With the TLT-supply the voltage and current are measured with the Tektronix high voltage probe (P6015) and a
Pearson current monitor (6585), respectively, at the locations indicated in Fig.~\ref{figsupplies}(b). For
all three supplies the signals are digitized on an oscilloscope using 0.2 ns sampling time (LeCroy
Waverunner 6100A).

The PM-supply at the electrical engineering department uses a differentiating-integrating system to measure
the fast high amplitude voltage and current waveforms \cite{hou90}. In this case a LeCroy Waverunner 2 is
used which has the same sampling time. The locations of the measurements are indicated in
Fig.~\ref{figsupplies}(c).

\subsection{Corona enclosure and electrodes}

Most measurements -- actually those with the C-supplies and the TLT-supply at the physics department -- were
carried out with the electrodes mounted in a large stainless steel vessel as drawn in Fig.~2
of~\cite{briscaling}. This vessel can be evacuated and filled with different gases with pressures in the
range of 0.013 to 1 bar, as thoroughly described in \cite{briscaling}. In the present paper, only
measurements at standard temperature and pressure in ambient air are presented, and a large window is taken
off the sidewall to ensure the refreshment of the air in the vessel. With the PM-supply at the electrical
engineering departement, that generates voltages up to 96 kV, the streamers propagate in the open air
without surrounding vacuum vessel. No significant differences were observed between streamers in open air
and in the stainless steel vessel when similar voltage pulses were applied.

All measurements presented in this paper are performed in a 40 mm gap with point-to-plane electrode
geometry. We have selected this gap length because in longer gaps, the streamers typically branch and
consecutively form thinner and slower streamers \cite{bri06}, while in a 40 mm gap, their diameter and
velocity is rather constant. In shorter gaps the streamers don't have enough space to develop fully.

The electrodes are as follows. The needle electrode is made from thoriated tungsten in C$_1$- and TLT-supply
or from pure tungsten in C$_2$- and PM-supply. No differences in streamer pattern are observed between these
two needles. The tip is spherical with radius $\approx$15 $\mu$m, a cone of height 2 mm connects the tip to
a cylinder with radius 0.5 mm; these shapes are not perfect, the surface has microscopic scratches from the
grinding process; even a dropped and misformed asymmetric needle does not create visibly different discharge
structures. Furthermore, for the C-supplies and the TLT-supply, the plate electrode is made from stainless
steel, while for the PM-supply, it is from aluminum.

\subsection{CCD-cameras and image data evaluation}\label{eval}

At the physics department, the discharges generated with the C-supplies and the TLT-supply are photographed
with a 4QuikE intensified CCD camera from Stanford Computer Optics as described in \cite{briscaling}. The
camera has 1360$\times$1024 pixels and a minimal exposure time of 2~ns. In the wave length range of 200 -
800~nm, it has a quantum efficiency of $\approx$0.3 on average. Each pixel has 12 bit, and different gains
can be chosen. The discharges generated with the PM-supply in the electrical engineering department are
photographed with a Princeton Instruments 576G/RB intensified CCD-camera as described in \cite{win06}. It
should be noted that if the initial streamer channels later evolve into arcs, the intensifier of the cameras
could be damaged even after exposure. Therefore a transition into an arc is carefully avoided by keeping the
voltage pulse sufficiently short.

A number of pictures are shown in our articles. However, it must be noted that the actual pictures depend on
the chosen representation of the raw data; this is demonstrated in panels a and b of Fig.~A2
of~\cite{briscaling}. An example of the actual data together with a picture is shown in Fig.~5
of~\cite{bri06}; profile bars of the real data along two sections through the image are shown on the margin.
Streamer diameters are evaluated as the full width at half maximum (FWHM) from this original data. The
further evaluation procedure is described in section 3.4.


\section{Measurements and results}\label{res}

\subsection{Streamer structures as function of voltage and polarity --- an overview}

\begin{figure}
\includegraphics[width=16cm]{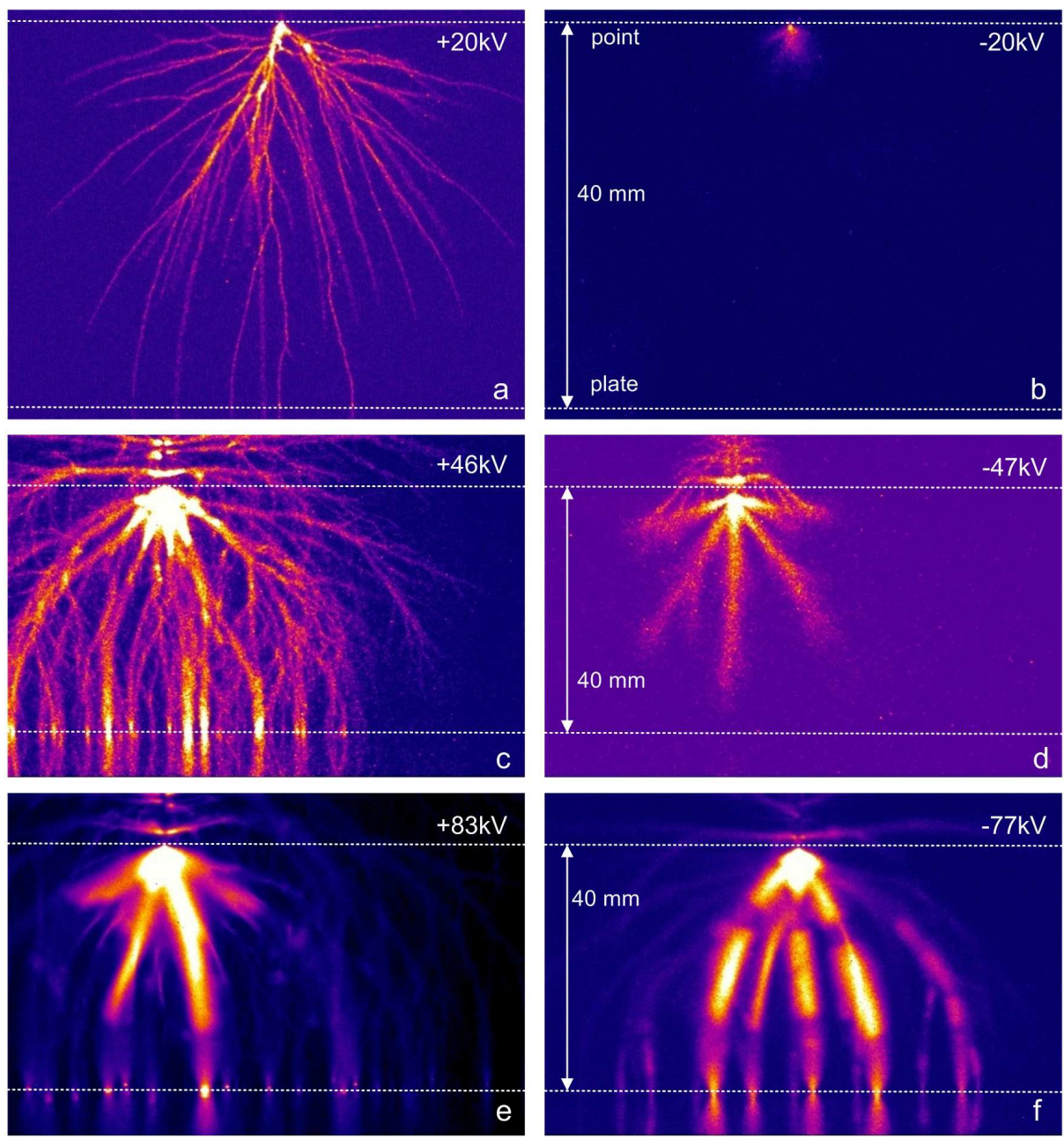}
\caption{Time integrated photographs of positive (left column) and negative (right column) streamers in a 40
mm gap in air at 1 bar. The applied voltages are indicated in the upper right corner of each panel, they are
$\approx\pm 20$~kV in the upper row, $\approx\pm 46$~kV in the middle row, and $\approx\pm 80$~kV in the
lower row. The discharges in panels a and b were made with the C$_2$-supply at the physics department, and
those in panels c - f with the PM-supply at the electrical engineering department; therefore different
cameras etc.\ were used. The white dotted lines indicate the position of the plate electrode below and the
height of the needle electrode tip above.} \label{posnegimpression}
\end{figure}

\begin{figure}
\centering
\includegraphics[width=12cm]{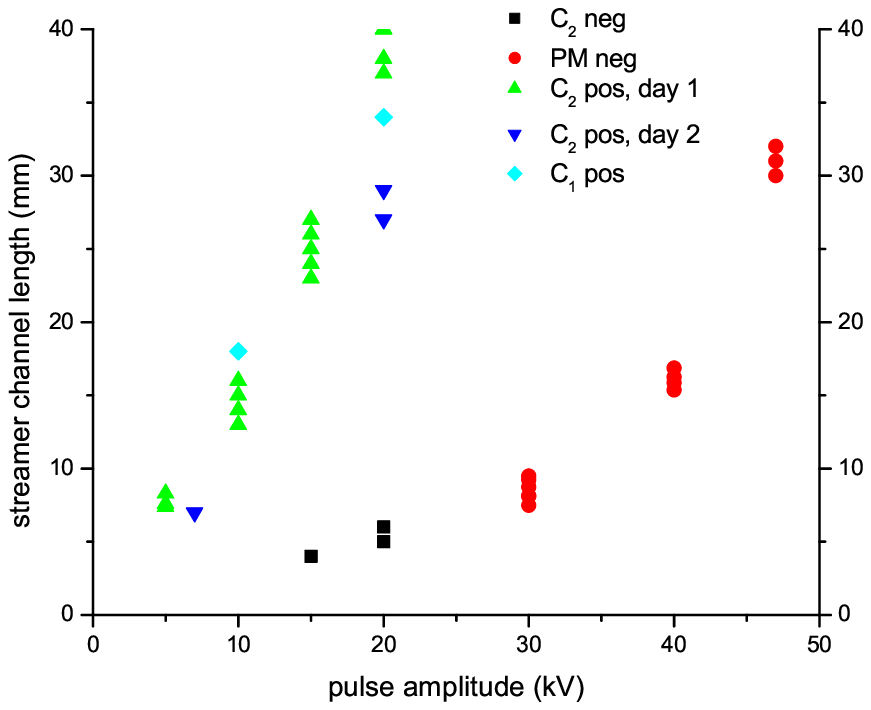}
\caption{Maximal streamer channel length as a function of voltage and polarity. The lengths are limited by
the distance between electrodes which is 40 mm. The longest streamers in each image are evaluated, they
typically move straight downward. Power supplies C$_1$, C$_2$ or PM are used as indicated in the figure. For
positive streamers created by the C$_2$-supply, the data for two different measurement days are shown, they
indicate a typical variation between days. Discharges below an extension of 10 mm are mostly cloud shaped
and will not be called streamers.} \label{lengths}
\end{figure}

As a first overview, Figure~\ref{posnegimpression} shows representative pictures of the discharges as a
function of voltage and polarity. The panels show the complete streamer evolution in a 40 mm point-plane gap
in air at 1 bar, i.e., with a sufficiently long exposure time of the photograph. In panels a and b, the slow
C$_2$-supply with the long pulse duration is used together with the equipment at the physics department. No
secondary streamers or return strokes appear after the primary streamers at voltages as low as 20~kV, and
the panels just show the primary discharge. In panels c to f, the PM-supply with a voltage pulse duration of
100~ns is used together with the equipment at the electrical engineering department; the images show the
streamers during the complete pulse duration. Streamers similar to those in panels c and d have also been
generated with the C$_1$- or the C$_2$-supply in the physics department; in particular, the temporal
evolution of positive discharges similar to panel c are described in detail in~\cite{bri06}. The initial
evolution of negative streamers similar to panel d is shown in Fig.~\ref{figdiag} below. The white dotted
lines indicate the approximate position of the plate electrode below and the height of the needle electrode
tip above. It can be seen that at high voltages, streamers are emitted from the complete needle electrode,
not only from its tip. Below the electrode spots in panels c, e and f, mirror images can be seen; they are
reflected from the planar electrode. The white lines do not precisely intersect with the foot points of all
streamers as the camera slightly looks down onto the electrode and the discharge is a three-dimensional
structure. This 3D structure can be resolved with stereographic imaging, for first results, we refer
to~\cite{nijdam}.

Figure~\ref{lengths} complements figure~\ref{posnegimpression}, it shows the maximal length of the streamer
channels as a function of voltage and polarity. The figures illustrate our statement that there are two
regimes with a smooth transition zone, a regime of voltages below 40 kV where positive and negative
discharges can be clearly distinguished, and a regime above 60 kV where they are quite similar. This will
now be elaborated in more detail.

Below 40~kV, positive and negative discharges are remarkably different. Positive discharges ignite above
5~kV, while negative discharges are only seen above 15~kV. When the voltage on the positive discharges
increases from 5 to 20 kV, thin streamers of increasing lengths form that branch frequently (for a further
characterization of branching, we refer to~\cite{bri06,briscaling,nijdam}); their lengths as a function of
voltage are shown in figure~\ref{lengths}. They bridge the gap above 20 kV as shown in
figure~\ref{posnegimpression}a. At this same voltage, the negative discharge only forms a small cloud around
the tip and no streamer, as can be seen in figure~\ref{posnegimpression}b\footnote{The positive streamers
also start from a cloud \cite{bri08,briscaling} although this cloud is not always visible, especially at
atmospheric pressure.}. Negative streamers form above 30~kV, but they are then still too short to determine
their diameters. Whenever their diameter can be determined, they are much thicker than the 0.2 mm diameter
observed for the thinnest positive streamers. Negative streamers bridge the complete 40 mm gap only at
voltages above 50~kV as shown in figure~\ref{lengths}. Panels c and d in figure~\ref{posnegimpression} show
positive and negative streamers at 46 kV where both are 1 to 2 mm thick but the negative ones do not reach
the plate electrode and hardly branch while the positive steamers cross the gap and form many branches. In
both cases, the discharge emerges not only from the needle tip, but also above it, most likely at sharp
edges.

Above 60~kV, positive and negative streamers become increasingly more alike. The negative streamers cross
the electrode gap at 56 kV (figure~\ref{figdiag}). At voltages above 75 kV, streamers of both polarities are
2 to 3 mm thick, bridge the gap and hardly branch. The positive streamers still branch a little more than
the negative ones (see panels e and f in figure~\ref{posnegimpression}).

We remark that figure~\ref{posnegimpression}f shows an overexposed streamer region in the middle of the gap
while the regions near the tip and near the plate are darker. The length of this region increases with
increasing voltage from 16 mm through 18 mm to 23 mm at 64 kV, 75 kV and 82 kV, respectively. It arises {\em
after} the primary streamers have bridged the electrode gap (where "primary streamer" denotes the first
group of streamers after the voltage has been applied~\cite{sig84}). It is not known why this large region
forms in the middle of the gap; usually such a region stretches out from the tip and is interpreted as a
secondary streamer or as a glow developing along the trails of the primary streamers (as in figure
\ref{posnegimpression}e). Striated or subdivided secondary streamers are also reported in~\cite{sig84},
however, without photographs. The spots on the plate electrodes also appear after the streamer has crossed
the gap; the cathode spots appear more or less spherical, whereas anode spots are contracted at the plate
but become broader and diffuser above the surface; they continue into the streamer more gradually.

\subsection{Dependence on the voltage supply}\label{DepSupply}

It should be noted that the streamers in figure~\ref{posnegimpression} are created with the C$_2$-supply at
20 kV and with the PM-supply at higher voltages. For positive streamers at voltages of at least 30~kV, we
have shown in~\cite{bri06} that the slow voltage rise time of the C$_2$-supply creates thinner channels than
the other supplies as the streamers start to propagate before the voltage has reached its maximal value;
this observation is also confirmed in the present measurements as demonstrated in panels b and c in
figure~\ref{posnegfotos}; these panels show positive streamer discharges at the same peak voltage of
$\approx$40~kV, but for voltage rise times of 30 or 150 ns, respectively. However, at 20~kV positive
streamers do not show this supply dependence as the streamer inception seems to be sufficiently slow at such
low voltages.

One might wonder whether a similar supply dependence holds for negative streamers, but we have no evidence
for it. Furthermore, one can speculate whether the negative streamers in figure~\ref{posnegimpression}(d) do
not cross the gap because the voltage pulse duration in the PM-supply is only 100 ns; if the streamer
inception time would be several tens of nanoseconds, the remaining pulse duration could be insufficient to
let the streamers cross the gap. To test these hypotheses, we have plotted the measured lengths of negative
streamers as a function voltage in figure~\ref{lengths}. The streamers at 15 and 20 kV are created with the
C$_2$-supply, and for voltages of 30 kV and larger with the PM-supply. The figure shows that the data follow
a smooth curve across the change of supplies; this suggests that the negative streamers are more robust
against changes of the power supplies than the positive ones, and that the negative streamers in
figure~\ref{posnegimpression} have reached their maximal length during the 100~ns duration of the voltage
pulse of the PM-supply.

\subsection{Stability field}\label{StabField}

A common hypothesis is that there is a stability field, i.e., a minimal field that the streamer needs to
propagate; this field would also be the characteristic field inside the streamer~\cite{rai91,all95}. This
hypothesis is guided by the empirical observation, that the ratio of applied voltage over maximal streamer
length is approximately constant for a given gas and polarity; this ratio has the dimension of an electric
field. (The role of some hypothetical average background field is discussed in the introduction.) For
positive streamers in point-plane and plane-plane gaps in air, a value of 4.4 to 5 kV/cm is fitted to
experiments in \cite{rai91,all95} and references therein. The data for positive streamers in
figure~\ref{lengths} is fitted by $U/L=6\pm1$~kV/cm, which is in agreement with the above values within the
error bar. For negative streamers, typical field values in the literature are a factor 2 or 3
higher~\cite{rai91,bab97}. Such values can also be fitted to our data for negative streamer in
figure~\ref{lengths}; however, it should be noted that the length is certainly not a linear function of the
voltage in this curve, also because there are no negative streamers below 40~kV, and that the approximation
of a constant stability field for negative streamers is questionable in view of this data.

\subsection{Streamer head diameters}

\begin{figure}
\includegraphics[width=16cm]{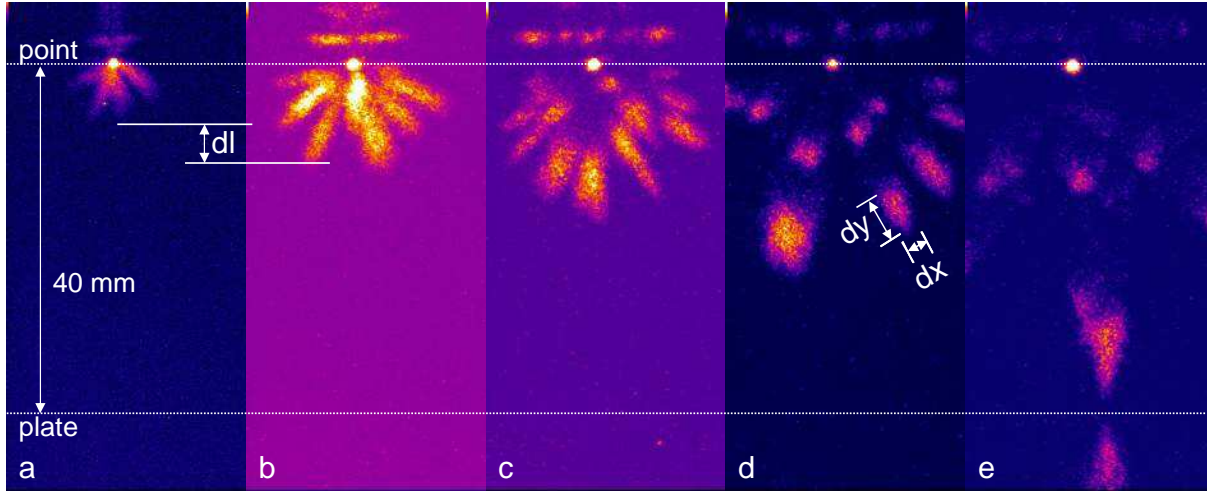}
\caption{Negative streamers at -56 kV generated with the PM-supply. The exposure time of each picture is
about 5 ns. The gate delay is (a) 4.4 ns, (b) 9.3 ns, (c) 14.3 ns, (d) 19.6 ns, (e) 28.8 ns. The white lines
indicate the position of the electrodes as in Fig.~\ref{posnegimpression}. The light at the bottom of figure
(e) is a reflection of the streamer light on the anode plate. How velocities are determined from such data,
is described in Section~\ref{vel}.} \label{figdiag}
\end{figure}

The streamer head diameters are determined from iCCD-images such as in figure \ref{figdiag}d; for the
evaluation of the raw data imaged in such figures, we refer to section~\ref{eval}. The full width $dx$ at
half maximum (FWHM) is measured and averaged over the $dy$-direction for as long as the streamer channel is
straight to suppress the stochastic single photon fluctuations within the pixels of the camera. The same
definition of the diameter was also used in \cite{bri06,bri08,briscaling}. Only single in-focus streamers at
a place without return stroke or electrode effects are evaluated, typically in the middle of the electrode
gap. Depending on voltage and polarity, typically three to ten streamers per photograph were evaluated, and
the diameters were furthermore averaged over three to five photographs per voltage. In the measurements with
the PM-supply, secondary streamers can appear after the primary streamers have crossed the gap, and
occasionally secondary streamers are evaluated instead of primary streamers when there are not sufficiently
many suitable primary streamer channels. However, primary or secondary streamer diameter show no significant
difference with the PM-supply with its high voltages, as also reported in \cite{hansieee,hansjpd}.

\begin{figure}
\centering
\includegraphics[width=10cm]{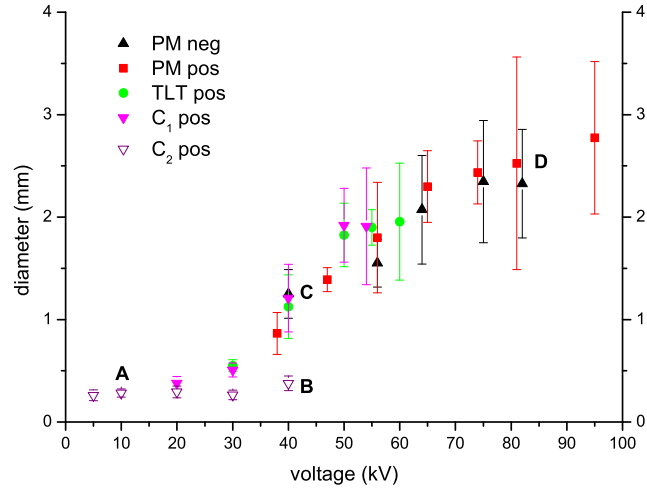}
\caption{Diameters of positive and negative streamers as a function of applied voltage. The symbols indicate
the different voltage supplies and polarities. The letters A, B, C, and D indicate where photographs of
positive streamers are included; they are shown in panels a, b, and c of figure~\ref{posnegfotos} for A, B
and C, and in figure \ref{posnegimpression}e for D.} \label{posnegdiam}
\end{figure}

\begin{figure}
\includegraphics[height=3.9cm]{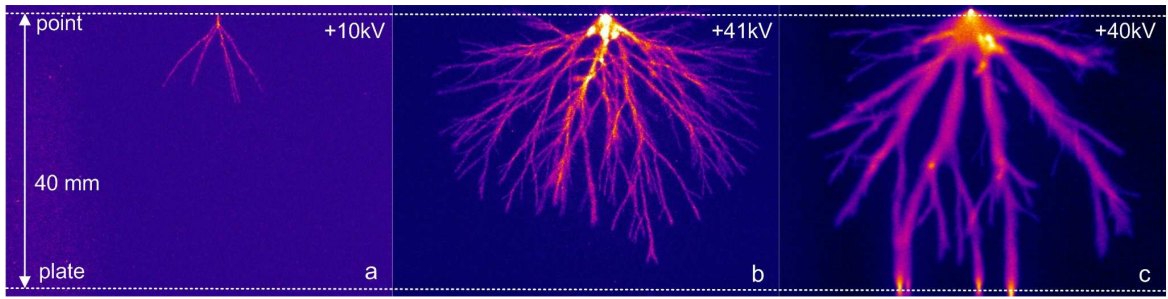}
\caption{Typical photographs of positive streamers for parameter values indicated by A, B, C in figure
\ref{posnegdiam}. Panels a and b are made with the slow C$_2$-supply, panel c with the fast C$_1$-supply.
The maximal voltage is a) +10~kV, b) +41~kV, and c) +40~kV. The white dotted lines indicate the electrode
positions as in Figs.~\ref{posnegimpression} and \ref{figdiag}. The streamers in panel a at 10~kV have
reached their maximal length; the streamers in panels b and c at $\approx$40~kV are still propagating.
Panels b and c demonstrate how strongly positive streamers at a voltage saturation value of 40~kV depend on
the voltage rise time of the supply, as discussed in~\cite{bri06}.} \label{posnegfotos}
\end{figure}

The diameters of positive and negative streamers as a function of applied voltage are shown in figure
\ref{posnegdiam}. The different power supplies and polarities are indicated with symbols. The measurements
of positive streamers with supplies C$_1$, TLT and PM together span a continuous curve within the error
bars, consistently with what we found in~\cite{bri06}. The positive streamers created with the C$_2$-supply
remain thin when the peak voltage increases, as is illustrated in panel b of figure~\ref{posnegfotos}; this
is also consistent with~\cite{bri06}. In this case, the streamers already start to propagate while the
voltage is still below its saturation value; furthermore their current is limited by the large series
resistance $R_3$. For negative streamers, diameters are only measured for experiments performed with the
PM-supply.

Positive streamers ignite above 5~kV. As shown in figure~\ref{posnegdiam}, their diameter then is 0.2 mm,
i.e., the minimal diameter \cite{bri06,briscaling}. For 20 kV, their diameter is 0.4 mm, and the streamers
bridge the 40 mm gap for the first time. Then the diameter increases by a factor of 10 (i.e., from 0.2 to 2
mm) when the applied voltage increases only by a factor of $\approx$2 (from 25 to 55 kV). Above 55 kV, the
diameter continues to increase, but less rapidly; it reaches 3 mm at 96 kV. Whether for higher voltages, the
diameter can become even higher, or whether it saturates eventually, is an open question. We observed that
at very high voltages, e.g. at +82 kV in figure \ref{posnegVI}a, the voltage rise time of the PM-supply
again becomes comparable to the streamer inception time as the inception time decreases with increasing
voltage; this entails that at high voltages the streamers start before the voltage has saturated. This is
also observed in \cite{hansjpd} with the same PM-supply at +74 kV, while it is not observed at -72 kV, in
agreement with the discussion in Section~\ref{DepSupply}.

The negative discharge becomes visible at 15 kV, but does not yet form streamers. The diameter of the
negative streamers can be measured at voltages of 40~kV or higher. For 40~kV the diameter is 1.2 mm. The
negative streamers bridge the gap for the first time at $\approx$ 56 kV (figure \ref{figdiag}) with a
diameter of 1.5 mm. Their diameter increases with applied voltage to $\approx$2.3 mm for voltages between 76
and 82 kV. On average, the positive streamers are about 10\% thicker than the negative ones, but this
difference is within the error margins.


\subsection{Streamer velocities}\label{vel}

\begin{figure}
\begin{minipage}[b]{.48\linewidth}
\includegraphics[height=6cm]{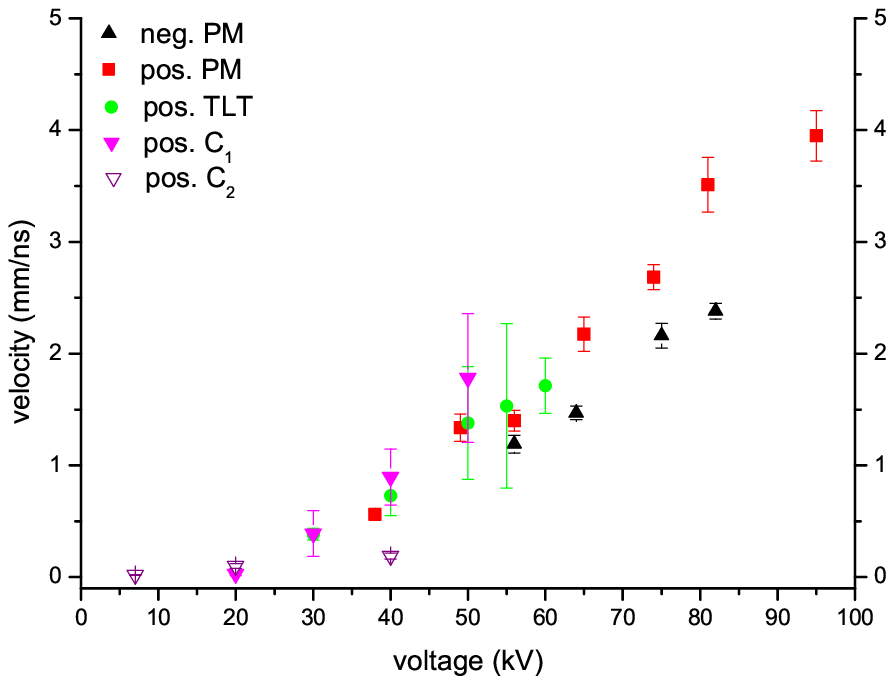}
\centering a)
\end{minipage}\hfill
\begin{minipage}[b]{.48\linewidth}
\includegraphics[height=6cm]{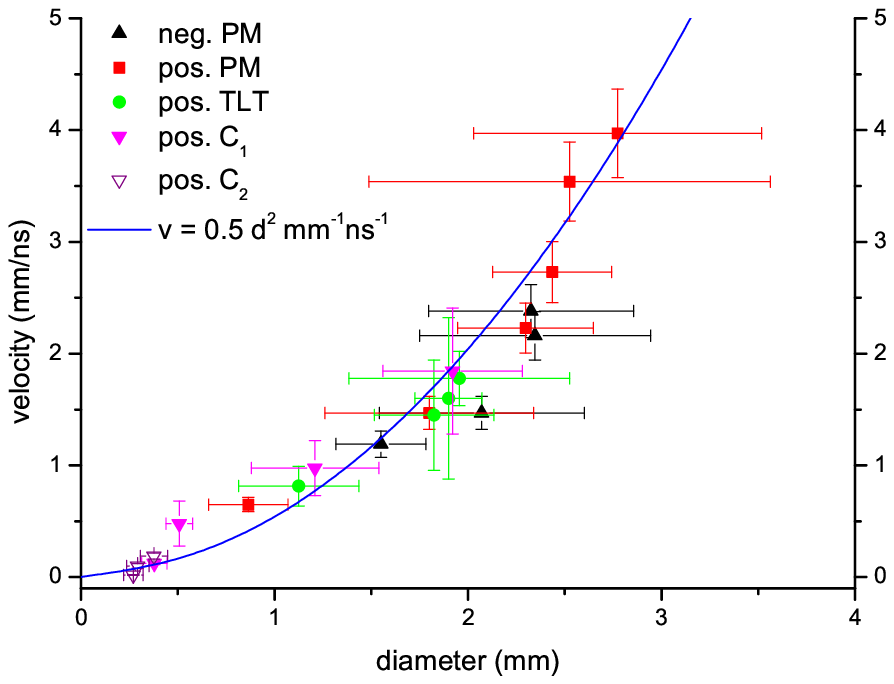}
\centering b)
\end{minipage}
\caption{a) Velocity of positive and negative streamers in air at standard temperature and pressure as a
function of voltage in a 4 cm gap. b) Velocity plotted as function of diameter. The line indicates a fit
with $v=0.5~d^2/$(mm~ns).} \label{posnegvel}
\end{figure}

In the experiments at the physics department with the C-supplies and the TLT-supply, the light trace on the
photographs can be interpreted as the path that the streamer head has followed during the exposure time of
the camera \cite{ebe06}; therefore the streamer velocities are measured as the full width at half maximum
(FWHM) of the length $dy$ in the propagation direction, divided by the exposure time $\Delta t$ of the
camera; the procedure is illustrated in figure~\ref{figdiag}. To minimize the error, sufficiently long
exposure times are used such that the streamer head crosses 1/4th to 1/8th of the electrode gap; these times
are 10 to 100 ns; the method is further described in \cite{briscaling}. The velocities are evaluated at some
distance from the electrodes; typically three to ten streamers per photograph and three to five photographs
per voltage are evaluated. In the presently investigated 40 mm gap, the streamer velocity and diameter are
approximately constant. In contrast, in the case of a 160 mm electrode gap the streamer velocity does depend
on the position and diminishes after branching~\cite{briscaling}.

In the experiments at the electrical engineering department with the PM-supply, the velocities are measured
as the propagation distance between photographs ($dl$ in figure \ref{figdiag}a,b) divided by the difference
in delay time $T$ of the photographs, where each photograph is taken during a different discharge pulse. The
streamers that have propagated furthest are evaluated in each picture, they predominantly move straight
downward. This method can be used since the jitter in streamer inception at high voltages is as small as
about 1 ns \cite{win07,hansieee}. Five to ten photographs with different delays (e.g., the five pictures in
figure \ref{figdiag}) are taken per voltage, depending on the gap transit time of the streamer. Position is
plotted as a function of time for each photograph and an average velocity is determined graphically. One or
two propagation sequences per voltage are evaluated. Only the fastest streamers are used since they
propagate most parallel to the camera's focal plane. Again the velocity is found to be constant throughout
the gap \cite{win07,hansieee}.

We remark that the ambiguity of the two-dimensional images of the three-dimensional discharge can lead to an
underestimation of the velocities that is counteracted by evaluating only the fastest streamers. Velocity
measurements with the stereographic imaging method introduced in~\cite{nijdam} are currently under way.

Figure \ref{posnegvel}a shows the resulting velocities of positive and negative streamers as a function of
voltage. The velocities of streamers of both polarities increase almost linearly with voltage. Velocities of
negative streamers at voltages lower than 56 kV could not be measured because the streamers did not
propagate far enough to measure them reliably. Note that the velocity of positive streamers increases
rapidly in this range, from 0.2 mm/ns at 20 kV to $\approx$ 1.5 mm/ns at 56 kV. Figure \ref{posnegvel}a
shows that the negative streamers are about 25\% slower than the positive ones when their velocities can be
measured.

\subsection{An empirical relation between diameter and velocity}\label{chdiamvel}

In figure \ref{posnegvel}b the velocity is plotted as a function of the diameter; the data is extracted from
figures \ref{posnegdiam} and \ref{posnegvel}a. The figure shows that the velocity $v$ increases with
increasing diameter $d$; it can be fitted quite well by the empirical equation $v = 0.5~ d^2/$(mm~ns) where
$v$ and $d$ are measured in their natural units. The approximation also works well for streamers of
intermediate radius as measured in~\cite{win07,hansieee,hansjpd}. However, streamers with a minimal diameter
of about 0.2~mm~\cite{bri06,briscaling} are a factor 5 too slow in this approximation. The velocities of the
thin streamers with diameters less than 0.5 mm in \cite{vel02,veld2003,panchpre} are also underestimated by
a factor of about 4.

\subsection{Current and energy}

\begin{figure}
\begin{minipage}[b]{.48\linewidth}
\includegraphics[width=8.3cm]{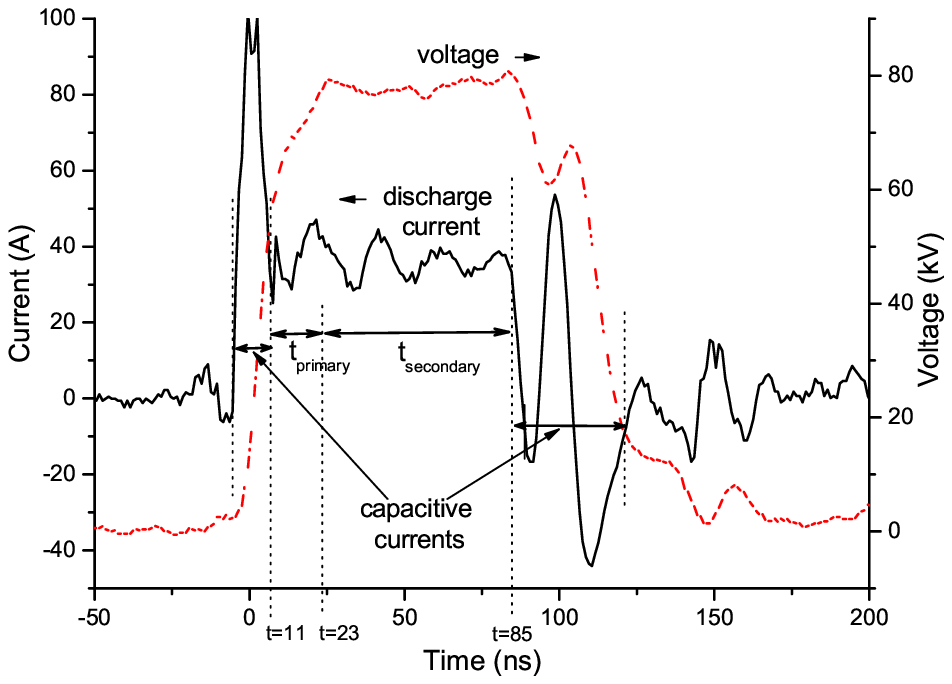}
\centering a)
\end{minipage}\hfill
\begin{minipage}[b]{.48\linewidth}
\includegraphics[width=8.3cm]{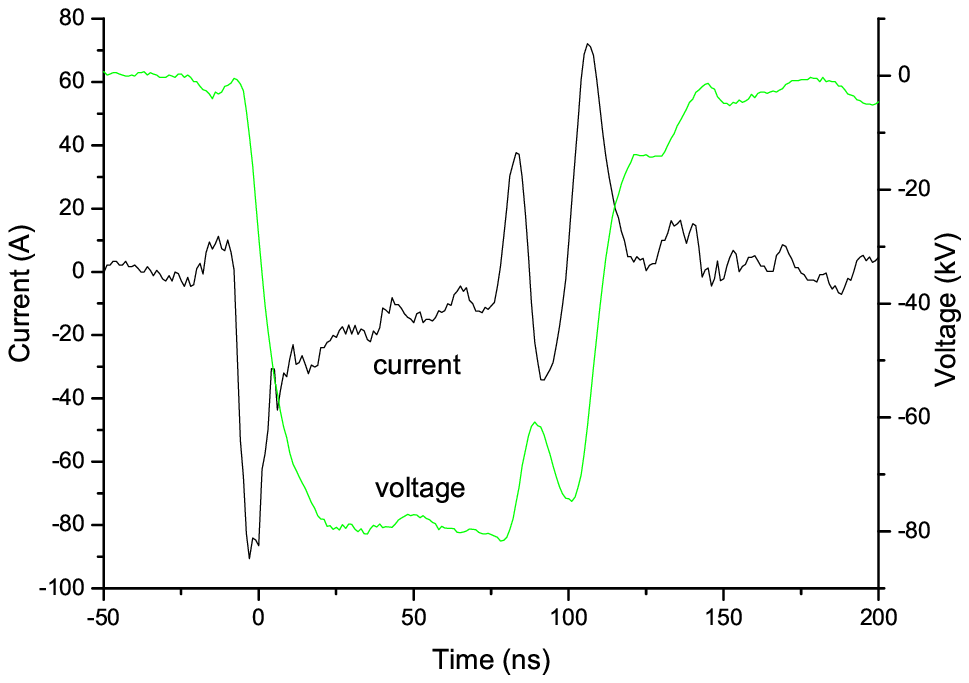}
\centering b)
\end{minipage}
\caption{Evolution of current and voltage for $\approx$ 82 kV for a) positive and b) negative streamers
generated with the PM-supply. The generated discharges are shown in panels e and f of
fig.~\ref{posnegimpression}. The current and voltage lines are smoothed by averaging over 5 adjacent points.
In panel a, characteristic times and capacitive versus discharge currents are indicated, as discussed in the
text.} \label{posnegVI}
\end{figure}

A typical evolution of current and voltage for positive and negative streamers with the PM-supply is shown
in Fig.~\ref{posnegVI}, the generated discharges are shown in panels e and f of Fig.~\ref{posnegimpression}.
The voltage pulse reaches approximately $\pm80$~kV and lasts for about 100~ns. While the voltage $U$ rises,
it creates a capacitive current $I_{capacitive}=C_{geom}~dU/dt$ in the circuit; $C_{geom}$ is the
capacitance of the discharge vessel in the absence of a discharge. The capacitive currents are indicated in
panel a of Fig.~\ref{posnegVI}. After approximately 11~ns (as indicated in the figure), the current is
mainly the discharge current; its maximum is approximately 45 A for both polarities. The positive discharge
current remains at this level for the duration of the voltage pulse (a plateau of duration $\approx$80 ns);
the oscillations are due to imperfect matching with the external circuit. The negative discharge current
does not have this plateau but slowly decreases to 10 A before the voltage pulse drops back to zero.

\begin{figure}
\centering
\includegraphics[width=11cm]{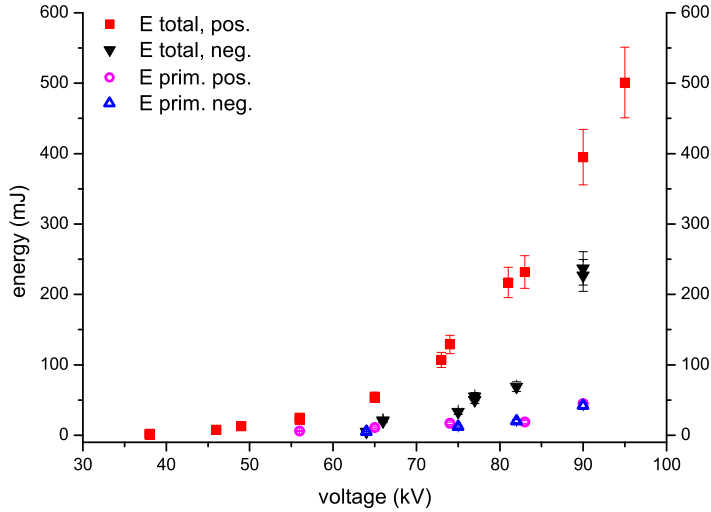}
\caption{Dissipated energy of the total discharge and of the primary positive and negative streamers only as
a function of voltage. The PM-supply was used.}
\label{posnegen}
\end{figure}

Estimates for the dissipated energies within the complete discharge are shown in figure~\ref{posnegen}; they
are calculated by integrating the product of voltage and current over time, after the capacitive part of the
total current is subtracted \cite{vel02}.
The total energy of the discharge is calculated by integrating to 200 ns when the voltage pulse certainly
has finished.

Fig.~\ref{figdiag} illustrates that the first group of streamers, the so-called primary
streamers~\cite{sig84}, cross the gap within 27~ns for $-56$~kV. This gap crossing time varies from 80 ns
for 30 kV to 10 ns for 96 kV for positive streamers, and from 27 ns for 56 kV to 15 ns for 83 kV for
negative ones. At a later time, secondary processes occur as described elsewhere, see
e.g.~\cite{win07,hansjpd,bri06}. The energy of the primary streamers is calculated by integrating from 11 ns
(when the total current becomes larger than the capacitive current, i.e., when the particle current begins
to flow) to the time when the primary streamers have bridged the gap. This arrival time is estimated from
the streamer velocity according to section~\ref{vel}. For 82~kV, it is 12~ns, therefore duration of the
primary streamers $t_{primary}$ is assumed to last from $t=11$~ns to $t=23$~ns, as indicated in panel a of
Fig.~\ref{posnegVI}.


Figure~\ref{posnegen} shows that the total energy of positive discharges is higher than for negative
discharges. This is because the current persists longer in the positive discharge while it decreases for the
negative ones (figure \ref{posnegVI}). The energy of the primary streamers is also shown in figure
\ref{posnegen}. Though these values must be regarded as an estimate, the data are similar for positive and
negative streamers. These energies are never higher than 50 mJ and thus are considerably lower than the
total energy which is maximally 400 mJ for a positive discharge and 220 mJ for a negative one at 90 kV. Our
observation that positive and negative streamers have similar energies, is consistent with
\cite{win07,hansieee,hansjpd}.


\section{Summary and conclusions}\label{concl}

Our measurements indicate that the inception and propagation processes of positive and negative streamers in
ambient air are quite different. Positive streamers emerge when a voltage of at least 5~kV is applied to a
needle electrode at 4 cm distance from a planar grounded electrode, while negative streamers only propagate
above 40~kV (which is around the DC-breakdown voltage of the gap). Two regimes have to be distinguished.
\begin{itemize}
\item[1)] $5~{\rm kV} < U < 40~{\rm kV}$: Positive streamers propagate. Their diameters increase from 0.2 mm
at 5 kV to 1 mm at 40 kV, they bridge the complete electrode gap above 20~kV and they branch. Their velocity
ranges from $10^5$~m/s at 5 kV to $10^6$~m/s at 40 kV. A negative discharge is only visible above 20 kV as a
glowing cloud near the electrode tip; no negative streamers are formed.
\item[2)] $U > 40~{\rm kV}$: Negative streamers appear, but they do not cross the gap for voltage below 56 kV.
Positive and negative streamers have a similar diameter of 1 mm at 40 kV that increases up to 3 mm for 96
kV. The positive streamer velocity increases from $10^6$~m/s at 40 kV to $4\cdot10^6$~m/s at 96 kV. The
velocity of negative streamers can only be measured for voltages higher than 56 kV which is then about 25\%
lower than for positive ones. The energy of the primary positive and negative streamers are similar; it
ranges from 20 to 50 mJ for voltages from 74 to 90 kV.
\end{itemize}
We also find the completely empirical fit $v = 0.5d^2/$(mm~ns) for the relation between velocity $v$ and
diameter $d$ of positive and negative streamers.

As a counterpart to the present experimental investigation of streamers of both polarities, a theoretical
investigation is presented in~\cite{PMTheo}. In that paper, the low voltage regime is investigated, and a
strong asymmetry between short positive and negative streamers is found, in agreement with the present
experiments. In particular, positive simulation streamers next to needle electrodes easily evolve out of
some initial ionization seed; they have small diameters comparable to experiments and a strong field
enhancement at the streamer tip; this field enhancement allows them to propagate into the regions with lower
background fields further away from the needle electrode. In these simulations, we find approximately the
same empirical relation between velocity and diameter for short positive streamers as in panel b of
Fig.~\ref{posnegvel}. On the other hand, negative simulation streamers evolve as well out of some initial
ionization seed in the high field region next to the electrode needle; but then their diameter increases
when penetrating the regions with lower fields; therefore the field is less enhanced and they easily
extinguish. We conclude that the simulations presented in~\cite{PMTheo} do allow to understand essential
physical features of the low voltage regime of the experiments presented here.

\ack
This work is financially supported by STW under contract number 06501 and by NWO under contract number 047.016.017.

\end{document}